\documentclass{lhep}       
\journal{LHEP}
\vol{xx}
\jyear{2018}
\pages{xxx} 
\usepackage{amsmath} 
\numberwithin{equation}{section}

\received{xx January 2018}
\published{xx March 2018}

\usepackage{mwe}
\usepackage{widetext}
\newcommand{\fR}{$f(R)$~}
\newcommand{\la}{\langle}
\newcommand{\ra}{\rangle}

\newcommand{\ie}{{\it i.e.}}

\begin{document}

\title{Quasi-Newtonian Cosmological Models in
Scalar-Tensor Theories of Gravity}
\author{Heba Sami and Amare Abebe }
\address{Center for Space Research, North-West University, South Africa\\
Department of Physics, North-West University, South Africa}

\begin{abstract}
In this contribution, classes of shear-free cosmological dust models with irrotational
fluid flows will be investigated in the context of scalar-tensor theories of gravity. In particular, the integrability conditions describing a consistent evolution of the linearised field equations of quasi-Newtonian universes are presented. We also derive the covariant density and velocity propagation equations of such models and analyse the corresponding solutions to these perturbation equations.
\end{abstract}

\maketitle


\section{{Introduction}}
Although general relativity theory (GR) is a generalization of Newtonian Gravity in the presence of strong gravitational fields, it has no properly defined  Newtonian limit  on cosmological scales.  Newtonian cosmologies are an extension of the Newtonian theory of gravity and are usually referred to as {\it quasi-Newtonian}, rather than strictly Newtonian formulations \cite{van1998quasi,maartens98,abebe2016integrability}. The importance of investigating the Newtonian limit for general relativity on cosmological contexts is that, there is a viewpoint that cosmological studies can be done using Newtonian physics, with the relativistic theory only needed for examination of some observational relations \cite{van1998quasi}. General relativistic quasi-Newtonian cosmologies have been studied in the context of large-scale structure formation and non-linear gravitational collapse in the late-time
universe. This despite the general covariant inconsistency of these cosmological models except in some special cases such as the spatially homogeneous and isotropic, spherically symmetric, expanding (FLRW) spacetimes. Higher-order or modified gravitational theories  of gravity such as $f(R)$  theories of gravity have been shown to exhibit more shared properties with Newtonian gravitation than does general relativity.  In \cite{van1998quasi}, a covariant approach to cold matter universes in quasi-Newtonian cosmologies has been developed and it has been applied and extended in \cite{maartens98} in order to derive and solve the equations governing density and velocity perturbations. This approach revealed the existence of integrability conditions in GR. In this work, we derive the evolution of the velocity and density perturbations in the comoving (Lagrangian) and quasi-Newtonian frames. We investigate the existence of integrability conditions of a class of irrotational and shear-free perfect fluid cosmological models in the context of scalar-tensor gravity. Such work has been done in the context of $f(R)$ gravity  \cite{gidelew2013beyond}, where some models of $f(R)$ gravity have been shown to exhibit Newtonian behaviour in the shear-free regime.

\subsection*{$f(R)$ and scalar-tensor models of gravitation}\label{fRST}
The so-called  $f(R)$ theories of gravity are among the simplest modification of Einstein's GR. These theories come about by a straightforward generalisation of the Lagrangian in the Einstein-Hilbert action \cite{sotiriou2010f, sotiriou2006f} as
\begin{eqnarray}\label{fRAct}
S_{f(R)}= \frac{1}{2}\int d^{4}x \sqrt{-g}\Big(f(R)+2\mathcal{L}_{m}\Big)\;,
\end{eqnarray}
where  $\mathcal{L}_{m}$ is the matter Lagrangian and $g$ is the determinant of the metric tensor $g_{\mu \nu}$.
Another modified theory of gravity is the scalar-tensor theory of gravitation. This is a broad class of gravitational models that tries to explain the gravitational interaction through both a scalar field and a tensor field. A sub-class of this theory, known as the  action the Brans-Dicke  (BD) theory,  has  an action of the form
\begin{eqnarray}\label{BDAct}
S_{BD}=\frac{1}{2}\int d^{4}x \sqrt{-g}\left[\phi R-\frac{\omega}{\phi}\nabla_{\mu}\phi\nabla^{\mu}\phi+2\mathcal{L}_{m}\right]\;,
\end{eqnarray} 
where $\phi$ is the scalar field and  $\omega$ is  a coupling constant considered to be independent of the scalar field $\phi$.
An interesting aspect of $f(R)$ theories of gravity is their proven equivalence with the BD theory of gravity \cite{sotiriou2006f,frolov2008singularity} with $\omega=0$. If we define the \fR  extra degree of freedom \footnote{$f'\;,f''$, etc. are the first, second, etc. derivatives of f w.r.t. the Ricci scalar R.} as
\begin{eqnarray}\label{fphi}
\phi\equiv f'-1\;,
\end{eqnarray}
then the actions \ref{fRAct} and \ref{BDAct} become dynamically equivalent.
In a FLRW background universe, the resulting non-trivial field equations lead to the following Raychaudhuri and Friedmann equations that govern the expansion history of the Universe \cite{carloni2008evolution}:
\begin{eqnarray}\label{6}
\dot{\Theta}+\frac{1}{3} \Theta^{2}= -\frac{1}{2(\phi+1)}\Big[ \mu_{m}+3p_{m}+f -R(\phi +1) + \Theta \dot{\phi} &\\ \nonumber + 3\phi^{''} \Big(\frac{\dot{\phi}^{2}}{\phi{'}^{2}}\Big)+ 3\ddot{\phi}- 3\frac{\dot{\phi} \dot{\phi}^{'}}{\phi^{'}}\Big]\;,
\end{eqnarray}
\begin{eqnarray}\label{7}
\Theta^{2}= \frac{3}{(\phi+1)}\Big[ \mu_{m} + \frac{R(\phi +1) - f}{2} + \Theta \dot{\phi}\Big]\;,
\end{eqnarray}
where $\Theta\equiv 3H = 3\frac{\dot{a}}{a}$, $H$ being the Hubble parameter, $a(t)$ is the scale factor, $\mu_{m}$ and $p_{m}$  are the energy density and isotropic pressure of standard matter, respectively.\\

The linearised thermodynamic quantities for the scalar field are the
energy density $\mu_{\phi}$, the  pressure $p_{\phi}$, the energy flux $q^{\phi}_{a}$ and the anisotropic pressure $\pi^{\phi}_{ab}$, respectively given by
\begin{eqnarray}\label{8}
&&\mu_{\phi}= \frac{1}{(\phi +1)} \Big[\frac{1}{2}\Big( R(\phi +1)-f\Big)- \Theta \dot{\phi} +  \tilde{\nabla}^{2}\phi \Big]\;,\\
&&\label{9} p_{\phi }= \frac{1}{(\phi +1)}\Big[ \frac{1}{2} \Big( f- R(\phi+1)\Big) + \ddot{\phi} - \frac{\dot{\phi} \dot{\phi}^{'}}{\phi^{'}}\\ &&
\nonumber + \frac{\phi^{''} \dot{\phi}^{2}}{\phi^{'2}}+ \frac{2}{3}(\Theta \dot{\phi} - \tilde{\nabla}^{2} \phi) \Big]\;,
\end{eqnarray}
\begin{eqnarray}\label{10}
&& q^{\phi}_{a}= -\frac{1}{(\phi+1)}\Big[ \frac{\dot{\phi}^{'}}{\phi^{'}} - \frac{1}{3} \Theta \Big]\tilde{\nabla}_{a} \phi\;,\\
&&\label{11}\pi ^{\phi}_{ab} = \frac{\phi^{'}}{(\phi+1)} \Big[\tilde{\nabla}_{\langle a} \tilde{\nabla}_{b \rangle}R - \sigma_{ab} \Big(\frac{\dot{\phi}}{\phi^{'}}\Big)\Big]\;.
\end{eqnarray}
The total ({\it effective})  energy density, isotropic pressure, anisotropic pressure and heat flux of standard matter and scalar field combination are given by
\begin{eqnarray}\nonumber
\mu \equiv\frac{\mu_{m}}{(\phi+1)}+ \mu_{\phi}, \hspace{.4cm} p \equiv \frac{p_{m}}{(\phi+1)}+ p_{\phi} \;,&&\\ \nonumber \pi_{ab}\equiv \frac{\pi^{m}_{ab}}{(\phi+1)}+ \pi^{\phi}_{ab}, \hspace{.4cm} q_{a}\equiv \frac{q^{m}_{a}}{(\phi+1)}+ q^{\phi}_{a}\;.
\end{eqnarray}
\subsection*{Covariant equations}\label{eq}
Given a choice of $4$-velocity field $u^{a}$ in the Ehlers-Ellis covariant approach \cite{ehlers1993contributions,ellis1989covariant}, the FLRW background is characterised by the equations  \cite{maartens98,abebe2016integrability}
\begin{eqnarray}\label{13}
&&\tilde{\nabla}_{a}\mu_{m}=0=\tilde{\nabla}_{a}p_{m}=\tilde{\nabla}_{a}\Theta\;,\\ &&  q^{m}_{a}=0=A_{a}=\omega_{a} \;,\\ &&\label{131} pi^{m}_{ab}= \sigma_{ab}=E_{ab}= 0=H_{ab}\;,
\end{eqnarray}
where $\Theta$, $A_{a}$, $\omega^{a}$, and  $\sigma_{ab}$ are the expansion, acceleration, vorticity and the shear terms.  $E_{ab}$ and $H_{ab}$ are the ``gravito-electric'' and ``gravito-magnetic'' components of the Weyl tensor $C_{abcd}$  defined from the Riemann tensor $R^a_{bcd} $ as 
\begin{eqnarray}\label{weyl}
&&C^{ab}{}_{cd}=R^{ab}{}_{cd}-2g^{[a}{}_{[c}R^{b]}{}_{d]}+\frac{R}{3}g^{[a}{}_{[c}g^{b]}{}_{d]}\;,\\
&&\label{geweyl}
E_{ab}\equiv C_{agbh}u^{g}u^{h},~~~~~~~H_{ab}\equiv\frac{1}{2}\eta_{ae}{}^{gh}C_{ghbd}u^{e}u^{d}\;.
\end{eqnarray}
The covariant linearised evolution equations in the general case are given by \cite{maartens98, abebe2016integrability}
\begin{eqnarray}\label{16}
&&\dot{\Theta}= -\frac{1}{3}\Theta^{2} - \frac{1}{2}(\mu+ 3p)+ \tilde{\nabla}_{a}A^{a}\;,\\
&&\label{17}
\dot{\mu}_{m}= -\mu_{m}\Theta -\tilde{\nabla}^{a}q^{m}_{a}\;,\\
&&\label{18}
\dot{q}^{m}_{a}= -\frac{4}{3}\Theta q^{m}_{a}- \mu_{m} A_{a}\;,\\
&&\label{19}
\dot{\omega}^{\langle a\rangle}= -\frac{2}{3}\Theta \omega^{a}- \frac{1}{2}\eta^{abc}\tilde{\nabla}_{b}A_{c}\;,\\
&&\label{20}
\dot{\sigma}_{ab}=-\frac{2}{3}\Theta \sigma_{ab}-E_{ab} + \frac{1}{2}\pi_{ab}+ \tilde{\nabla}_{\langle a}A_{b\rangle}\;,\\
&&\label{21}
\dot{E}^{\langle ab\rangle}= \eta^{cd\langle a}\tilde{\nabla}_{c} H^{\rangle b}_{d}- \Theta E^{ab} - \frac{1}{2}\dot{\pi}^{ab}-\frac{1}{2}\tilde{\nabla}^{\langle a}q^{b\rangle}\\ \nonumber &&- \frac{1}{6}\Theta \pi^{ab}\;,\\
&&\label{22}
\dot{H}^{\langle ab\rangle}= -\Theta H^{ab}- \eta^{cd\langle a}\tilde{\nabla}_{c}E^{\rangle b}_{d}+ \frac{1}{2} \eta^{cd\langle a}\tilde{\nabla}_{c}\pi^{\rangle b}_{d}\;,
\end{eqnarray}
and the linearised constraint equations are given by
\begin{equation}\label{23}
C^{ab}_{0}\equiv E^{ab}- \tilde{\nabla}^{\langle a}A^{b \rangle}- \frac{1}{2} \pi^{ab}=0\;,
\end{equation}
\begin{equation}\label{24}
C^{a}_{1}\equiv \tilde{\nabla}_{b}\sigma^{ab}- \eta^{abc}\tilde{\nabla}_{b}\omega_{c}- \frac{2}{3}\tilde{\nabla}^{a}\Theta+ q^{a}=0\;,
\end{equation}
\begin{equation}\label{25}
C_{2}\equiv \tilde{\nabla}^{a}\omega_{a}=0\;,
\end{equation}
\begin{equation}\label{26}
C^{ab}_{3}\equiv\eta_{cd(}\tilde{\nabla}^{c}\sigma^{d}_{b)}+\tilde{\nabla}^{\langle a}\omega^{b\rangle}-H^{ab}=0\;,
\end{equation}
\begin{equation}\label{27}
C^{a}_{5}\equiv\tilde{\nabla}_{b}E^{ab}+\frac{1}{2}\tilde{\nabla}_{b}\pi^{ab}-\frac{1}{3}\tilde{\nabla}^{a}\eta+\frac{1}{3}\Theta q^{a}=0\;,
\end{equation}
\begin{equation}\label{28}
C^{a}_{b}\equiv\tilde{\nabla}_{b}H^{ab}+(\mu +p)\omega^{a}+\frac{1}{2}\eta^{abc}\tilde{\nabla}_{b}q_{a}=0\;.
\end{equation}
\section{{Quasi-Newtonian spacetimes}}
If a comoving  $4$-velocity $\tilde{u}^{a}$ is chosen such that, in the linearised form 
\begin{equation}\label{29}
\tilde{u}^{a}= u^{a}+v^{a}, \hspace{.3cm} v_{a}u^{a}=0, \hspace{.3cm} v_{a}v^{a}<<1\;,
\end{equation}
the  dynamics, kinematics and gravito-electromagnetics quantities \ref{13}-\ref{131} undergo transformation.\\
Here $v^{a}$ is the relative velocity  of the comoving frame with respect to the  observers in the quasi-Newtonian frame, defined such that it vanishes in the background. In other words, it is a non-relativistic peculiar velocity.  Quasi-Newtonian cosmological models are irrotational, shear-free dust spacetimes characterised by \cite{maartens98, abebe2016integrability}:
\begin{equation}\label{30}
 p_{m}=0\;, \hspace*{.3cm} q^{m}_{a}=\mu_{m} v_{a}\;, \hspace*{.3cm} \pi^{m}_{ab}=0\;,\omega_{a}=0\;, \hspace*{.3cm}  \sigma_{ab}=0\;.
\end{equation}
The gravito-magnetic constraint Eq. \ref{26} and the shear-free and irrotational condition \ref{30} show that the gravito-magnetic component of the Weyl tensor automatically vanishes:
\begin{equation}\label{32}
H^{ab}=0\;.
\end{equation}
The vanishing of this quantity implies  no  gravitational radiation in quasi-Newtonian cosmologies, and Eq. \ref{28} together with Eq. \ref{30} show that $q^{m}_{a}$ is irrotational and thus so is $v_{a}$:
\begin{equation}
\eta^{abc}\tilde{\nabla}_{b}q_{a}=0 = \eta^{abc}\tilde{\nabla}_{b}v_{a}\;.
\end{equation}
Since the vorticity vanishes, there exists a velocity potential such that
\begin{equation}\label{33}
v_{a}= \tilde{\nabla}_{a}\Phi\;.
\end{equation}
\section{{Integrability conditions}}\label{eqq}
It has been shown that the non-linear models are generally inconsistent if the silent constraint \ref{32} is imposed, but that the linear models are consistent \cite{maartens98, abebe2016integrability}. Thus, a simple approach to the integrability conditions for quasi-Newtonian cosmologies follows from showing that these models are in fact a sub-class of the linearised silent models. This can  happen by using the transformation between the quasi-Newtonian and comoving frames.\\
The transformed linearised kinematics, dynamics and gravito-electromagnetic quantities from the quasi-Newtonian frame to the comoving frame are given as follows:
\begin{equation}\label{34}
\tilde{\Theta} =\Theta +\tilde{\nabla}^{a}v_{a}\;,
\end{equation}
\begin{equation}\label{35}
\tilde{A}_{a}= A_{a}+ \dot{v}_{a}+\frac{1}{3}\Theta v_{a}\;,
\end{equation}
\begin{equation}\label{36}
\tilde{\omega}_{a}= \omega_{a}- \frac{1}{2}\eta_{abc}\tilde{\nabla}^{b}v^{c}\;,
\end{equation}
\begin{equation}\label{37}
\tilde{\sigma}_{ab}= \sigma_{ab}+ \tilde{\nabla}_{\langle a}v_{b\rangle}\;,
\end{equation}
\begin{equation}\label{38}
\tilde{\mu}= \mu, \hspace{.3cm} \tilde{p}=p, \hspace{.3cm} \tilde{\pi}_{ab}= \pi_{ab}, \hspace{.3cm} \tilde{q}^{\phi}_{a}= q^{\phi}_{a}\;
\end{equation}
\begin{equation}\label{39}
\tilde{q}^{m}_{a}= q^{m}_{a}-(\mu_{m} +p_{m})v_{a}\;,
\end{equation}
\begin{equation}\label{40}
\tilde{E}_{ab}= E_{ab}, \hspace{.3cm} \tilde{H}_{ab}= H_{ab}\;.
\end{equation}
It follows from the above transformation equations that
\begin{eqnarray}\label{41}
&&\tilde{p}_m=0\;,\hspace*{.3cm} \tilde{q}^m_{a}=0=\tilde{A}_{a}=\tilde{\omega}_{a}\;, \\ \nonumber && \tilde{\pi}^m_{ab}=0=\tilde{H}_{ab}\;,\hspace*{.3cm}
\tilde{\sigma}_{ab}=\tilde{\nabla}_{\langle a}v_{b\rangle}\;,\hspace*{.3cm}\tilde{E}_{ab}= E_{ab}\;.
\end{eqnarray}
These equations describe the linearised silent universe except that the restriction on the shear in Eq. \ref{41} results in the integrability conditions for the quasi-Newtonian models. Due to the vanishing of the shear in the quasi-Newtonian frame, Eq. \ref{20} is turned into a new constraint 
\begin{equation}\label{44}
E_{ab}- \frac{1}{2}\pi^{\phi}_{ab}-\tilde{\nabla}_{\langle a}A_{b\rangle}=0\;.
\end{equation}
This can be simplified by using Eq. \ref{19} and the identity for any scalar $\varphi$: 
\begin{equation}\label{45}
\eta^{abc} \tilde{\nabla}_{a}A_{c}=0 \Rightarrow A_{a}= \tilde{\nabla}_{a}\varphi\;.
\end{equation}
In this case $\varphi$ is the covariant relativistic generalisation of the Newtonian potential.
\subsection*{First integrability condition}
Since Eq. \ref{44} is a new constraint, we need to ensure its consistent propagation at all epochs and in all spatial hypersurfaces. Differentiating it with respect to cosmic time $t$ and by using equations \ref{11}, \ref{21} and \ref{24}, one obtains
\begin{eqnarray}\label{46}
\tilde{\nabla}_{\langle a}\tilde{\nabla}_{b\rangle}\Big[ \dot{\varphi} +\frac{1}{3} \Theta +\frac{\dot{\phi}}{(\phi+1)}\Big]+\Big[\dot{\varphi}+\frac{1}{3} \Theta \\ \nonumber +\frac{\dot{\phi}}{(\phi+1)}\Big]\tilde{\nabla}_{a}\tilde{\nabla}_{b}\varphi =0\;,
\end{eqnarray}
which is the first integrability condition for quasi-Newtonian cosmologies in scalar-tensor theory of gravitation and it is a generalisation of the one obtained in \cite{maartens98}, {\it i.e.}, Eq.\ref{46} reduces to an identity for the generalized van Elst-Ellis condition  \cite{van1998quasi,maartens98,abebe2016integrability}
\begin{equation}\label{47}
\dot{\varphi}+\frac{1}{3}\Theta= -\frac{\dot{\phi}}{(\phi+1)}\;.
\end{equation}
The evolution equation of the $4$-acceleration $A_{a}$ can be shown, using Eqs. \ref{47} and \ref{24}, to be
\begin{eqnarray}\label{49}
\dot{A}_{a}+\Big[\frac{2}{3}\Theta+\frac{\dot{\phi}}{(1+\phi)}\Big]A_{a}= -\frac{1}{2(1+\phi)}\Big[ \mu_{m} v_{a}+ \Big(\frac{1}{3}\Theta \\ \nonumber + \frac{\dot{\phi}^{\prime}}{\phi^{\prime}}- \frac{2\dot{\phi}}{(1+\phi)}\Big)\tilde{\nabla}_{a}\phi\Big]\;.
\end{eqnarray}
\subsection*{Second integrability condition}
There is a second integrability condition arising by checking for the consistency of the constraint \ref{44} on any spatial hyper-surface of constant time $t$. By taking the divergence of \ref{44} and by using the following identity:
\begin{equation}\label{50}
\tilde{\nabla}^{b}\tilde{\nabla}_{\langle a}A_{b\rangle}= \frac{1}{2}\tilde{\nabla}^{2}A_{a}+ \frac{1}{6}\tilde{\nabla}_{a}(\tilde{\nabla}^{c}A_{c})+\frac{1}{3}(\mu -\frac{1}{3}\Theta^{2})A_{a}\;,
\end{equation}
which holds for any projected vector $A_{a}$, and by using Eq. \ref{45}  it follows that:
\begin{equation}\label{51}
\tilde{\nabla}^{b}\tilde{\nabla}_{\la a}\tilde{\nabla}_{b\ra}\varphi= \frac{2}{3}\tilde{\nabla}_{a}(\tilde{\nabla}^{2}\varphi)+\frac{2}{3}(\mu-\frac{1}{3}\Theta^{2})\tilde{\nabla}_{a}\varphi\;.
\end{equation}
 By using Eqs. \ref{51}, \ref{24} and \ref{27}, one obtains:
\begin{eqnarray}\label{52}
&\tilde{\nabla}_{a}\mu_{m}- \Big[ \dot{\phi}+\frac{2}{3} (\phi+1)\Theta\Big]\tilde{\nabla}_{a}\Theta+ \frac{1}{(\phi+1)}\Big[ \frac{f}{2}-\mu_{m}\\ \nonumber &+\Theta\dot{\phi}  - \frac{\Theta \dot{\phi}^{\prime}(\phi+1)}{\phi^{\prime}}\Big]\tilde{\nabla}_{a}\phi -2(\phi+1)\tilde{\nabla}^{2}(\tilde{\nabla}_{a}\varphi) -2\Big[\mu_{m}\\ \nonumber &+ \frac{R(\phi+1)}{2} -\frac{f}{2}-\Theta\dot{\phi}-\frac{\Theta^{2}(\phi+1)}{3}\Big]\tilde{\nabla}_{a}\varphi  -\tilde{\nabla}^{2}(\tilde{\nabla}_{a}\phi)=0\;,
\end{eqnarray}
which is the second integrability condition and in general it appears to be independent of the first integrability condition \ref{46}.
By taking the gradient of Eq. \ref{47} and using Eq. \ref{24}, one can obtain the peculiar velocity:
\begin{equation}\label{53}
v_{a}= -\frac{1}{\mu_{m}}\Big[ 2(\phi+1)\tilde{\nabla}_{a}\dot{\varphi}+ \Big( \frac{\dot{\phi}^{\prime}}{\phi^{\prime}}-\dot{\varphi}-\frac{3\dot{\phi}}{(\phi+1)}\Big)\tilde{\nabla}_{a}\phi\Big]\;.
\end{equation} 
By virtue of Eqs. \ref{17} and \ref{18}, $v_{a}$ evolves according to
\begin{equation}\label{54}
\dot{v_{a}}+\frac{1}{3}\Theta v_{a}= -A_{a}\;.
\end{equation}
The coupled evolution Eqs. \ref{49} and \ref{54} decouple to produce the second-order propagation equation of the peculiar velocity $v_{a}$. By using Eqs. \ref{6} and \ref{7} in Eq. \ref{54} one obtains:
\begin{eqnarray}\label{55}
\ddot{v}_{a}+\Big[\Theta+\frac{\dot{\phi}}{(\phi+1)}\Big]\dot{v}_{a}+\Big[ \frac{1}{9} \Theta^{2} -\frac{1}{6(\phi+1)}(5\mu_{m} -f \\ \nonumber -4\Theta \dot{\phi})\Big]v_{a}
+ \frac{1}{(\phi+1)}\Big[ \frac{\dot{\phi}}{(\phi+1)}- \frac{\phi^{''}}{2\phi^{'}}- \frac{\Theta}{6} - \frac{\dot{\phi}^{'}}{2\phi^{'}}\\ \nonumber  + \frac{\phi^{''} \dot{\phi}}{2\phi^{'2}}\Big]\tilde{\nabla}_{a}\phi=0\;.
\end{eqnarray}
\section{{Perturbations}}
In the previous section, we showed how imposing special restrictions to the linearized perturbations of FLRW universes in the quasi-Newtonian setting  result in the integrability conditions. These integrability conditions imply velocity and acceleration propagation equations resulting from the  generalised van Elst-Ellis condition for the acceleration potential in scalar-tensor theories. In this section, we show how one can obtain the velocity and density perturbations via these propagation equations, thus generalizing GR results obtained in \cite{maartens98}.\\
\newline
We define the variables that characterise scalar inhomogeneities the matter
energy density, expansion, peculiar velocity, acceleration as well as the scalar fluid and scalar field momentum, respectively, as follows:
\begin{eqnarray}\label{Delta}
&&\Delta_{m}= \frac{a^{2} \tilde{\nabla}^{2}\mu_{m}}{\mu_{m}}\; ,\\
&&\label{Z}
Z= a^{2}\tilde{\nabla}^{2}\Theta \; ,\\
&&\label{V}
V^{m}=a^{2}\tilde{\nabla}^{a}v_{a}\; ,\\
&&\label{Ascalar}
\mathcal{A}= a^{2}\tilde{\nabla}^{a}A_{a}\; ,\\
&&\label{phiscalar}
\Phi= a^{2}\tilde{\nabla}^{2}\phi\; \\
&&\label{psiscalar}
\Psi= a^{2}\tilde{\nabla}^{2}\dot{\phi}\; .
\end{eqnarray}
The system of equations governing the evolutions of these scalar fluctuations are given as follows

\begin{eqnarray}\label{Vddot}
&&\ddot{V}^{m}+\left( \frac{\Theta}{3}+ \frac{\dot{\phi}}{(1+\phi)}\right)\dot{V}^{m}- \frac{1}{2(\phi+1)}\Big[ \mu_{m} V^{m} \\ \nonumber &&+ \Big( \frac{1}{3} \Theta  + \frac{\dot{\phi}^{'}}{\phi^{'}}-\frac{2\dot{\phi}}{(\phi+1)}\Big)\Phi\Big]=0\; ,\\
&&\label{Deltaddot}
\ddot{D}^{m}+ \left[\frac{\dot{\phi}}{(\phi+1)}\right]\dot{D}^{m}- \Big[\frac{3\mu_{m}}{2(\phi+1)}\Big]D^{m}- \Theta \ddot{V}^{m} \\ \nonumber &&+\Big[\frac{\Theta^{2}}{3}+ \frac{5\mu _{m}}{2(\phi+1)}- \frac{f}{2(\phi+1)}+ \frac{\tilde{\nabla}^{2}\phi}{(\phi+1)}\\ \nonumber && + \frac{3\dot{\phi}\dot{\phi}^{'}}{2\phi^{'}(\phi+1)}- \frac{3\Theta \dot{\phi}}{2(\phi+1)}\Big]\dot{V}^{m}-\tilde{\nabla}^{2}\dot{V}^{m} \\ \nonumber &&+ \left(\frac{2}{3} \Theta  -\frac{\dot{\phi}}{(\phi+1)}\right)\tilde{\nabla}^{2}V^{m}
+ \frac{\Theta}{(\phi+1)}\dot{\Phi}\\ \nonumber &&+ \frac{1}{2(\phi+1)}\Big[ \frac{2f}{(\phi+1)} -\frac{3\ddot{\phi}^{'}}{\phi^{'}}- \frac{4\mu_{m}}{(\phi+1)} +\frac{2}{3}\Theta^{2} \\ \nonumber &&+ \frac{4\Theta \dot{\phi}}{(\phi+1)} - \frac{\tilde{\nabla}^{2}\phi}{(\phi+1)} - \frac{7\Theta \dot{\phi}^{'}}{\phi^{'}}\Big]\Phi=0\; ,\\
&&\label{phiscalarddot}
\ddot{\Phi}- \frac{\dot{\phi}^{'}}{\phi^{'}} \dot{\Phi}- \Big[\frac{\ddot{\phi}^{'}}{\phi^{'}} \frac{\dot{\phi}^{'2}}{\phi^{'2}} -\frac{\Theta \dot{\phi}}{6(\phi+1)} -\frac{\dot{\phi}\dot{\phi}^{'}}{2\phi^{'}(\phi+1)}\\ \nonumber && + \frac{\dot{\phi}^{2}}{(\phi+1)^{2}}\Big]\Phi -\Big[ \frac{\Theta \dot{\phi}}{3}+ \frac{\dot{\phi}^{2}}{(\phi+1)}-\ddot{\phi}\Big]\dot{V}^{m}\\ \nonumber && + \frac{\dot{\phi} \mu_{m}}{2(\phi+1)}V^{m}=0\; .
\end{eqnarray}
\subsection*{Harmonic decomposition}
Since the evolution equations obtained so far are too complicated to be solved, the harmonic decomposition approach is applied to these equations using the eigenfunctions and the corresponding wave number for these equations, therefore we write
\begin{equation}
X= \sum_{k}X^{k}Q_{k}(\vec{x})\; , \hspace*{1cm} Y=\sum_{k}Y^{k}(t)Q_{k}(\vec{x})\;,
\end{equation}
where $Q_{k}(x)$ are the eigenfunctions of the covariantly defined spatial Laplace-Beltrami
operator\cite{gidelew2013beyond,carloni2006gauge}, such that
\begin{equation}
\tilde{\nabla}^{2}Q= -\frac{k^{2}}{a^{2}}Q\; .
\end{equation}
The order of the harmonic (wave number) is given by
\begin{equation}
k=\frac{2\pi a}{\lambda}\; ,
\end{equation}
where $\lambda$ is the physical wavelength of the mode. The eigenfunctions $Q$ are covariantly constant, \ie
\begin{equation}
\dot{Q}_{k}(\vec{x})=0\; .
\end{equation}
Applying the harmonic decomposition, the second-order evolution equations \ref{Vddot}-\ref{phiscalarddot} can be rewritten as
\begin{eqnarray}\label{VHddot}
&&\ddot{V}^{k}_{m}+\left( \frac{\Theta}{3}+ \frac{\dot{\phi}}{(1+\phi)}\right)\dot{V}^{k}_{m} -\frac{1}{2(\phi+1)}\Big[ \mu_{m} V^{k}_{m} \\ \nonumber &&+ \left( \frac{1}{3} \Theta + \frac{\dot{\phi}^{'}}{\phi^{'}}-\frac{2\dot{\phi}}{(\phi+1)}\right)\Phi^{k}\Big]=0\; ,\\
&&\label{DeltaHddot}
\ddot{\Delta}^{k}_{m}+ \frac{\dot{\phi}}{(\phi+1)}\dot{\Delta}^{m} - \frac{3\mu_{m}}{2(\phi+1)} \Delta^{k}_{m}- \Theta \ddot{V}^{k}_{m}
\\ \nonumber &&+\Big[ \frac{\Theta^{2}}{3}+ \frac{5\mu _{m}}{2(\phi+1)}- \frac{f}{2(\phi+1)}- \frac{k^{2}\phi}{a^{2}(\phi+1)}\\ \nonumber &&  + \frac{3\dot{\phi}\dot{\phi}^{'}}{2\phi^{'}(\phi+1)}- \frac{3\Theta \dot{\phi}}{2(\phi+1)}+\frac{k^{2}}{a^{2}}\Big]\dot{V}^{k}_{m}\\ \nonumber && -\Big[\frac{2k^{2}}{3a^{2}} \Theta  -\frac{\dot{\phi}k^{2}}{(\phi+1)a^{2}}\Big]V^{k}_{m}
+ \frac{\Theta}{(\phi+1)}\dot{\Phi}^{k} \\ \nonumber &&+ \frac{1}{2(\phi+1)}\Big[\frac{2f}{(\phi+1)} -\frac{3\ddot{\phi}^{'}}{\phi^{'}}- \frac{4\mu_{m}}{(\phi+1)} +\frac{2}{3}\Theta^{2} \\ \nonumber &&+ \frac{4\Theta \dot{\phi}}{(\phi+1)} - \frac{\tilde{\nabla}^{2}\phi}{(\phi+1)} - \frac{7\Theta \dot{\phi}^{'}}{\phi^{'}}\Big]\Phi^{k}=0\; ,\\
&&\label{phiscalarddot}
\ddot{\Phi}^{k}- \frac{\dot{\phi}^{'}}{\phi^{'}} \dot{\Phi}^{k}- \Big[ \frac{\ddot{\phi}^{'}}{\phi^{'}}- \frac{\dot{\phi}^{'2}}{\phi^{'2}} -\frac{\Theta \dot{\phi}}{6(\phi+1)} \\ \nonumber &&-\frac{\dot{\phi}\dot{\phi}^{'}}{2\phi^{'}(\phi+1)} + \frac{\dot{\phi}^{2}}{(\phi+1)^{2}}\Big]\Phi^{k} -\Big( \frac{\Theta \dot{\phi}}{3} \\ \nonumber &&+ \frac{\dot{\phi}^{2}}{(\phi+1)}-\ddot{\phi}\Big)\dot{V}^{k}_{m} + \frac{\dot{\phi} \mu_{m}}{2(\phi+1)}V^{k}_{m}=0\;.
\end{eqnarray}
\section{{Solutions}}
In this section, we consider $R^{n}$ model, one of the $f(R)$ toy models that are considered to be the simplest and widely studied form of higher order $f(R)$ gravitational theories.\\
The Lagrangian density of such models is given as
\begin{equation}
f(R)= \beta R^{n}\;,
\end{equation} 
where $\beta$ represents the coupling parameter and an arbitrary constant $n\neq1$ is considered for exploring cosmological models. In \cite{carloni2005cosmological}, it has been shown, using the cosmological dynamical systems approach, that the scale factor $a(t)$ admits an exact solution of the form
\begin{equation}
a= a_{0}t^{\frac{2n}{3(1+w)}}\;,
\end{equation}
with $w=0$ and normalized coefficients $\beta$ and $a_{0}$. One can obtain the following expressions for the expansion, the Ricci scalar and the effective matter energy density respectively:
\begin{eqnarray}
&&\Theta= \frac{2n}{t}\; , \hspace*{.3cm} R=\frac{4n(4n-3)}{3t^{2}}\; ,\\
&&
\mu_{m}= n\Big(\frac{3}{4}\Big)^{1-n} \Big(\frac{4n^{2}-3n}{t^{2}}\Big)^{n-1}\Big(\frac{-16n^{2}+26n-6}{3t^{2}}\Big)\;.
\end{eqnarray}
Therefore we have the perturbation equations \ref{VHddot}, \ref{DeltaHddot} and \ref{phiscalarddot} as
\small \begin{eqnarray}\label{VHddot1}
&&\ddot{V}^{k}_{m}+ \Big(\frac{(6-4n)}{3t}\Big)\dot{V}^{k}_{m}- \left(\frac{8n^{2}-13n+3}{3t^{2}}\right)V^{k}_{m} \\ \nonumber &&- \Big[frac{4t^{2n-3}}{3\left(\frac{4n(4n-3)}{3}\right)^{n-1}}\Big]\Phi_{k}=0\;,\\
&&\label{DeltaHddot1}
\ddot{\Delta}^{k}_{m}+\left[\frac{2(1-n)}{t}\right]\dot{\Delta}^{k}_{m}+\left(\frac{3+13n-8n^{2}}{t^{2}}\right)\Delta^{k}_{m} \\ \nonumber &&-\left(\frac{2n}{t}\right)\ddot{V}^{k}_{m}+ \Big[ \frac{(62n^{2}-127n+27)}{3t^{2}}+\frac{k^{2}t^{\frac{2(n-3)}{3}}}{n\Big(\frac{4n(4n-3)}{3}\Big)^{n-1}}  \\ \nonumber &&+6n^{2}-6n\Big]\dot{V}^{k}_{m}- \left[ \frac{2k^{2}(3-n)}{3t^{\frac{(4n+3)}{3}}}\right]V^{k}_{m}+ \left[\frac{2t^{(2n-3)}}{\Big( \frac{4n(4n-3)}{3}\Big)^{n-1}}\right]\dot{\Phi}_{a}\\ \nonumber &&+\frac{t^{2n}}{n\Big(\frac{4n(4n-3)}{3}\Big)^{n-1}}\Big[ \frac{(28n^{2}-8n)}{3t^{4}} - \frac{6(2n^{2}-7n+6)}{t^{3}} \\ \nonumber &&+ \frac{k^{2}}{t^{\frac{(4n+6)}{3}}}- \frac{k^{2}}{nt^{\frac{(12-2n)}{3}}n\Big(\frac{4n(4n-3)}{3}\Big)^{n-1}}\Big]\Phi_{k}=0\; ,\\
&&\label{P}
\ddot{\Phi}_{k}-\left(\frac{(4-2n)}{t}\right)\dot{\Phi}_{k}-\left(\frac{(8n^{2}-8n+12)}{3t^{2}}\right)\Phi_{k}\\ \nonumber &&-
(-2n+2)\left(\frac{4n(4n-3)}{3}\right)^{n-1}\left(\frac{2n^{2}}{3t^{2}}+2n^{2}-3n+2\right)t^{-2n}\dot{V}^{k}_{m}\\ \nonumber &&+ n(1-n)\left(\frac{4}{3}\right)^{1-n}\left(\frac{4n^{2}-3n}{t^{2}}\right)^{n-1}\left(\frac{16n^{2}+26n-6}{3t^{3}}\right)V^{k}_{m}=0\; . 
\end{eqnarray}
\subsection*{Exact solutions}
In this subsection, we will solve the perturbations equations we obtained so far. The exact solutions of the density and velocity perturbation equations are found in the comoving frame, using the $f(R)$ solutions in \cite{carloni2008evolution} and a simple workaround. A simple alternative is then  to work in the comoving frame and apply the transformation from the comoving frame to the quasi-Newtonian frame using the following identity \cite{maartens98} \\
\begin{equation}\label{transformation}
\tilde{D}_{a}f= \tilde{\nabla}_{a}f+\dot{f}v_{a}\; .
\end{equation}
The comoving perturbation variables are defined as
\begin{eqnarray}\label{tildeD}
&&\tilde{\Delta}^{m}_{a}= \dfrac{a\tilde{D}_{a}\mu_{m}}{\mu_{m}}\; ,\\
&&\label{tildeZ}
\tilde{Z}_{a}= a\tilde{D}_{a}\Theta\; ,\\
&&\label{tildephi}
\tilde{\Phi}_{a}=a\tilde{D}_{a}\phi\; ,\\
&&\label{tildepsi}
\tilde{\Psi}_{a}= a\tilde{D}_{a}\dot{\phi}\; .
\end{eqnarray}
By using the identity \ref{transformation}, the comoving perturbation variables can be rewritten as
\begin{eqnarray}\label{tildeD1}
&&\tilde{\Delta}^{m}_{a}= \Delta^{m}_{a}- \Theta V^{m}_{a}\; ,\\
&&\label{tildeZ1}
\tilde{Z}_{a}= Z_{a}-\Big[\dfrac{1}{3}\Theta^{2}+\dfrac{1}{2(\phi+1)}\Big(2\mu_{m}-f-2\Theta \dot{\phi}\\ \nonumber && +2\tilde{\nabla}^{2}\phi\Big)V^{m}_{a}\Big]\;  ,\\
&&\label{tildephi1}
\tilde{\Phi}_{a}=\Phi_{a}+\dot{\phi}V^{m}_{a}\; ,\\
&&\label{tildepsi1}
\tilde{\Psi}_{a}= \Psi_{a}+\ddot{\phi}V^{m}_{a}\; .
\end{eqnarray}
The second-order evolution equation of the density perturbation in the comoving frame admits a general solution of the form \cite{carloni2008evolution}
\begin{equation}\label{Deltasolcomoving}
\tilde{\Delta}^{k}_{m}= C_{1}t^{-1}+C_{2}t^{\alpha_{+}}+C_{3}t^{\alpha_{-}}-C_{4} C_{0}t^{2-\dfrac{4n}{3}}\; ,
\end{equation}
where $C_{1}$, $C_{2}$, $C_{3}$ and $C_{4}$ are constants, $\alpha_{\pm}$ is given as
\begin{equation}\label{alpha}
\alpha_{\pm}= -\dfrac{1}{2}\pm \dfrac{\sqrt{(n-1)(256n^{3}-608n^{2}+417n-81)}}{6(n-1)}\; ,
\end{equation}
and $C_{0}$ is the conserved value for the gradient variable $C_{a}$, where
\begin{equation}\label{Ca}
C_{a}= a^{3} \tilde{\nabla}_{a}\tilde{R}\; ,
\end{equation}
where $\tilde{R}$ is the three dimension Ricci scalar, is defined as
$$\tilde{R}= 2\mu -\dfrac{2}{3}\Theta^{2}\; .$$ \\ 
Therefore, by using Eqs. \ref{tildeD1} and \ref{Deltasolcomoving}, the general solution of the density perturbation Eq. \ref{DeltaHddot} in the quasi-Newtonian frame can be written as
\begin{equation}\label{Delatsolquasi}
\Delta^{k}_{m}= C_{1}t^{-1}+C_{2}t^{\alpha_{+}}+C_{3}t^{\alpha_{-}}-C_{4}C_{0}t^{2-\dfrac{4n}{3}}+\dfrac{2n}{t} V^{k}_{m}\;.
\end{equation}
The gradient variable $\Phi_{a}$ is equivalent to $\mathcal{R}_{a}$ defined in $f(R)$ theory \cite{gidelew2013beyond}, such that
\begin{equation}
\Phi_{a}= \phi^{'} \mathcal{R}_{a}\; . \label{phiRA}
\end{equation}
The solution of $\mathcal{R}$ has been obtained in the comoving frame \cite{carloni2008evolution} and it has the form 
\begin{equation}
\mathcal{R}= C_{5}t^{-3}+ C_{6}t^{\beta_{+}}+C_{7}t^{\beta_{-}}-C_{8}C_{0}t^{-\dfrac{4n}{3}}\; .
\end{equation}
Therefore, the solution of the second-order perturbation Eq. \ref{P} in the comoving frame can be written as 
\begin{equation}\label{Phisolcomoving}
\tilde{\Phi}_{k}= n(n-1)\left( \dfrac{4n(4n-3)}{3t^{2}}\right)^{n-2}\left[C_{5}t^{-3}+ C_{6}t^{\beta_{+}}+C_{7}t^{\beta_{-}}-C_{8}C_{0}t^{-\dfrac{4n}{3}}\right]\; ,
\end{equation}
Therefore, by using Eq. \ref{tildephi1}, the general solution of the perturbation Eq. \ref{P} in the quasi-Newtonian frame is given as 
\begin{eqnarray}\label{Phiquasisol}
&&\Phi_{k}= n(n-1)\left(\dfrac{4n(4n-3)}{3t^{2}}\right)^{n-2}\Big(C_{5}t^{-3}+ C_{6}t^{\beta_{+}}+C_{7}t^{\beta_{-}}\\ \nonumber &&-C_{8}C_{0}t^{-\dfrac{4n}{3}}+ \left(\dfrac{8n(4n-3)}{3t^{3}}\right)V^{k}_{m}\Big)\;.
\end{eqnarray}
Using Eq. \ref{Phiquasisol}, the general solution of the perturbation Eq. \ref{VHddot} in the quasi-Newtonian frame is given as
\begin{equation}\label{Vquasisol}
V^{k}_{m}= \left( K_{2}+ K_{3}(B_{1}-B_{2})\right)t^{\dfrac{2n}{3}+\gamma_{-}}+\left( K_{1}+ K_{3}(B_{3}+B_{4})\right)t^{\dfrac{2n}{3}+\gamma_{+}}\; ,
\end{equation}
where $K_{1}$, $K_{2}$ are two arbitrary constants and 
\begin{eqnarray}
&& K_{3}= \dfrac{3(n-1)\left(\dfrac{16n^{2}}{3}-4n\right)^{n}}{(4n-3)\sqrt{-8n^{2}-48n(2n+2)+132n-27}}\; ,\\
&& \gamma_{\pm}=-\dfrac{1}{2}\pm \dfrac{\sqrt{-80n^{2}-48n(2n+2)+132n-27}}{6}\; ,\\
&& B_{1}= C_{1}\int t^{\left(-\dfrac{2n}{3}-\dfrac{3}{2}\right)+\gamma_{+}}dt\; ,\\
&&B_{3}= C_{1}\int t^{\left(-\dfrac{2n}{3}-\dfrac{3}{2}\right)+\gamma_{-}}dt\; ,
\end{eqnarray}
where $B_{2}$ and $B_{4}$ are all functions of time and their expressions are rather complicated.\\
Thus, we get the full set of exact solutions for the density and velocity perturbation equations in the quasi-Newtonian frame.

\subsection{Approximate solutions}
In this subsection, we apply a quasi-static approximation to the evolution equations \ref{DeltaHddot} and \ref{VHddot}. 
In this approximation, terms involving time derivatives for gravitational potential are neglected and only  those terms involving density perturbation are kept \cite{starobinsky2007disappearing,tsujikawa2008observational,gidelew2013beyond}. In \cite{gidelew2013beyond}, a quasi-static approximation for the matter perturbation has been introduced for both radiation and dust dominated epochs. A quasi-static approximation was taken such that the time evolutions of $\mathcal{R}_{a}$ are neglected,  $\dot{\mathcal{R}}_{a} \simeq 0$ and $\ddot{\mathcal{R}}_{a} \simeq 0$. 

According to Eq. \ref{phiRA}, the time variations in $\Phi_{a}$ are neglected, $\ie $ $\dot{\Phi}_{a} \simeq 0$ and $\ddot{\Phi}_{a} \simeq 0$.\\
Hence, one can get
\begin{equation}\label{VHddot2}
\ddot{V}^{k}_{m}+ \left(\dfrac{48n^{2}-108n+7n}{(36-18n)t}\right)\dot{V}^{k}_{m}+\left(\dfrac{(8n^{2}-13n+3)}{3t^{2}}\right)V^{k}_{m}=0\; .
\end{equation}
This second-order equation admits a general solution of the form
\begin{equation}\label{VHddotsolutionquasi}
V^{k}_{m}(t)= C_{9}t^{D+E_{+}}+C_{10}t^{D+E_{-}}\; ,
\end{equation}
where
$$D= \dfrac{\sqrt{-32n^{4}+300n^{3}-723n^{2}+588n-108}}{6(n-2)}\; ,$$
and 
$$E_{\pm}= \pm\dfrac{(8n^{2}-15n+6)}{6(n-2)}\; .$$ 
Based on this solution in Eq. \ref{VHddotsolutionquasi} and its first and second time derivative, the general solution of Eq. \ref{DeltaHddot1} is 
\begin{equation}\label{Deltaddotsolutionquasiapprox}
\Delta(t)= C_{11}t^{-\dfrac{1}{2}+n+L_{+}}+ C_{12}t^{-\dfrac{1}{2}+n+L_{-}}\; ,
\end{equation}
where $C_{11}$ and $C_{12}$ are arbitrary constants and 
$$L_{\pm}= \pm \dfrac{\sqrt{36n^{2}-56n-11}}{2}\; .$$
There are some other solutions to Eq. \ref{DeltaHddot1} which are rather too complicated to be presented here.
\hfill
\newline

\underline{\it Conclusion}~:~
\newline
Our main goal has been the study of the cosmological perturbation in the context of one of the modified theories of gravity. We reviewed two of these alternative theories of gravity,
namely $f(R)$ and scalar-tensor theories. We investigated classes of shear-free cosmological dust models with irrotational fluid flows in the context of scalar-tensor theories. We presented the integrability conditions that describe a consistent evolution of the linearised field equations of quasi-Newtonian universes. We defined the gradient variables that characterize the cosmological perturbations and derived the second-order evolution equations of these variables.  The harmonic decomposition approach is applied to these equations in order to solve this complicated system of differential equations. After getting a complete set of the perturbation equations, we solved these equations by considering $R^{n}$ models to get the exact solutions for the density and velocity perturbations. We introduced the so-called quasi-static approximation to admit the approximated solutions on small scales. Solving the whole system numerically has been left for future work.

\section*{Acknowledgements}
HS acknowledges the financial support of the Center for Space Research, North-West University and the  Centre of Excellence-Mathematical and Statistical Sciences (CoE-MaSS), Wits University to attend "Beyond Standard Model" conference of the Center for Fundamental Physics (CFP)
Zewail City of Science and Technology. AA acknowledges that this work is based on the research supported in part by the National Research Foundation of South Africa, the Faculty Research Committee of the Faculty of Natural and Agricultural Sciences of North-West University and the Centre of Excellence in  Mathematical and Statistical Sciences (CoE-MaSS), University of the Witwatersrand.

\bibliographystyle{unsrt}

\end{document}